\documentclass[pdflatex,sn-basic]{sn-jnl}
\usepackage{graphicx}
\graphicspath{{figures/}}
\usepackage{amsmath,amssymb,amsfonts}
\usepackage{booktabs}
\usepackage[section]{placeins} 
\usepackage{hyperref}
\usepackage{xcolor}
\newcommand{\SPI}[1]{\ensuremath{\mathrm{SPI}\mbox{-}#1}}
\newcommand{\SMRI}[1]{\ensuremath{\mathrm{SMRI}\mbox{-}#1}}

\begin{document}
\title[Snow, drought characterization, and drought forecasting in Ohio]{Snow Changes Drought Characterization but Not Drought Predictability: A Leakage-Free Multiscale Study in Ohio}
\author[1]{\fnm{Symon} \sur{Forthman}}
\author[1]{\fnm{Max } \sur{Gluck}}
\author*[1,2]{\fnm{Fatma Sevin\c{c}} \sur{Kurnaz}}\email{fskurnaz@yildiz.edu.tr; fatma.kurnaz@case.edu}
\affil[1]{\orgname{Case Western Reserve University}, \orgaddress{\city{Cleveland}, \state{Ohio}, \country{USA}}}
\affil[2]{\orgname{Y\i ld\i z Technical University}, \orgaddress{\city{Istanbul}, \country{Turkey}}}
\abstract{We examine whether accounting for snow accumulation and melt, which precipitation-based drought indices ignore, changes drought characterization and one-month-ahead drought forecasting at three stations in Ohio, USA (1953--2026). The Standardized Precipitation Index (SPI) is compared with the Standardized Snow Melt and Rain Index (SMRI), and both are decomposed with the Haar maximal overlap discrete wavelet transform. Nonlinear autoregressive networks without (NAR) and with (NARX) wavelet inputs are compared over 100 paired repetitions in a leakage-free design (past-only wavelet inputs, untouched test period), together with persistence, climatology, and linear autoregressive benchmarks. Snow changes the 3-month index mainly from December to May (mean absolute SMRI--SPI differences of up to 0.17--0.28). These differences are attributable to the snow-adjusted input rather than to the fitted distribution, are robust to the snow-model parameters, and change the drought classification of 21--32\% of the December--May drought months. Wavelet inputs reduce NARX error relative to NAR by 6--10\% for the 3-month indices in 92--100\% of the repetitions, but not for the 6-month indices. A linear autoregression with 7--12 lags is at least as accurate as two-lag NARX for every series, and networks given six lags gain at most 4.5\% from the wavelet inputs and only match the autoregression. The forecast gain therefore reflects the memory of the accumulated index rather than multiscale information, and it is similar for the SPI and the SMRI. Snow matters for drought monitoring in winter and spring, but not for one-month predictability.}
\maketitle

\section{Introduction}
\label{sec:intro}

Drought is one of the costliest and least predictable natural hazards. Unlike floods or storms, it develops slowly, has no clear onset or end, and its impacts spread through agriculture, water supply, ecosystems, and the economy long after precipitation has returned to normal. Because the same rainfall deficit can matter in very different ways for different sectors, there is no single definition of drought. It is usually classified as meteorological, agricultural, hydrological, or socioeconomic, depending on the part of the water cycle and the type of impact considered \citep{wilhite1985,mishra2010,vanloon2015}. Humid regions are not immune. In 2012, a rapidly intensifying ``flash'' drought struck the central United States and caused severe agricultural losses \citep{otkin2016}. In 2024, Ohio experienced its most intense drought in 25 years, and parts of the state reached the Exceptional (D4) category of the U.S.\ Drought Monitor for the first time since the monitor began in 2000 \citep{osu2025}. Such events show the value of reliable drought monitoring and of forecasts that can support early warning, even in regions that are generally considered water-rich.

Drought is commonly monitored with standardized indices. The most widely used is the Standardized Precipitation Index \citep[SPI;][]{mckee1993}. The SPI fits a probability distribution to precipitation accumulated over a chosen number of months and transforms it to a standard normal variable. Because the accumulation period can be varied, the same index describes short-term conditions relevant to soils and crops as well as long-term conditions relevant to streamflow and groundwater. The SPI requires only precipitation data, and the World Meteorological Organization has recommended it to all national meteorological and hydrological services \citep{svoboda2012}.

The SPI, however, treats all precipitation as if it were immediately available. In regions with seasonal snow cover, this is not the case. Precipitation that falls as snow is stored in the snowpack and released later as melt, sometimes weeks or months later. A precipitation-based index can therefore indicate normal conditions while little water is actually reaching the ground, and vice versa. This is the basis of the concept of \emph{snow drought}, in which snow storage is anomalously low either because precipitation is low (``dry'' snow drought) or because warm temperatures turn snow into rain or cause early melt (``warm'' snow drought) \citep{harpold2017,huning2020}. Several indices have been proposed to account for these processes. Among them, the Standardized Snow Melt and Rain Index \citep[SMRI;][]{staudinger2014} keeps the SPI methodology but replaces precipitation with the simulated daily sum of rainfall and snowmelt, and it requires only daily precipitation and temperature. \citet{staudinger2014} showed for Swiss catchments that the SMRI captured snow-related drought information that the SPI missed. Other approaches build snow storage into physically based moisture-anomaly indices \citep{zhang2019szi,tian2022}. Most applications of snow-aware indices, however, concern mountain or high-latitude regions where snowmelt dominates the water supply. Fewer studies consider regions where snow is seasonal and variable rather than dominant.

Ohio is such a region. Snow is not the main water source, but it is a recurrent part of the winter water balance. In the Ohio River basin, snow ablation events peak in late winter and early spring; about 40\% of them involve rain on snow, and their frequency has declined in the northern and eastern parts of the basin \citep{suriano2024}. Across the eastern United States, warm winters have become much more frequent since the 1960s and are associated with less snowfall, shallower snowpacks, more bare-ground days, and earlier peak streamflow \citep{ford2021}. In such a climate, the timing of water input to the land surface can differ noticeably from the timing of precipitation in some months while hardly differing in others. It is therefore an open question whether a snow-aware drought representation changes the information available for drought monitoring and, in particular, for drought forecasting.

Drought forecasting methods range from statistical and dynamical models to data-driven approaches \citep{mishra2011,hao2018,fung2020}. Among data-driven tools, artificial neural networks (ANNs) are attractive because they can represent nonlinear relationships without a prescribed model form, and they have been used to forecast drought indices for almost two decades \citep{mishra2006}. Dynamic architectures are particularly suited to time series. The nonlinear autoregressive (NAR) network predicts an index from its own past values, and the NAR network with exogenous inputs (NARX) also uses past values of other series. NAR-type networks have been applied, for example, to forecast drought indices in hyper-arid Kuwait \citep{alsumaiei2020}.

A large part of the recent literature combines such models with wavelet decomposition. The wavelet transform separates a time series into components that describe its variability at different time scales, and these components are then used as model inputs. Since early applications to drought forecasting \citep{kim2003}, wavelet-based hybrids have been combined with neural networks, support vector regression, fuzzy logic, genetic programming, and extreme learning machines, and they have often been reported to outperform the corresponding single models \citep{ozger2012,belayneh2014,mehr2014,deo2017}. The approach is reviewed by \citet{nourani2014} and remains an active area of research \citep[e.g.][]{tugrul2025}. In this line of work, \citet{evkaya2021} forecast the SPI at the Bursa station, Turkey, with NAR and NARX networks. Haar wavelet series were selected according to their Kendall rank correlation with the SPI, and the original SPI was combined with these series in a NARX model. Using the original index together with its wavelet decompositions in a NARX model improved forecasting performance compared with NAR models.

Two methodological issues have since become clear in the wavelet-forecasting literature, and both are relevant here. The first is the use of future information. A wavelet decomposition, and in particular a multiresolution analysis, computes the value of a component at time $t$ from observations on both sides of $t$. When the whole record is decomposed before it is split into training and test periods, the inputs used to ``forecast'' a test value therefore contain information that would not exist at the time of the forecast. \citet{quilty2018} showed that this widely overlooked property of the wavelet transform had led to a large number of incorrectly developed wavelet-based forecasting models that cannot be used for real-world forecasting, and they proposed best practices for building models that use only information available at the forecast date \citep[see also][]{du2017}. The second issue concerns how wavelet components are selected. Significance tests of the association between an index and its components assume independent observations and, with long monthly records, can declare almost every component significant. Such tests then say little about which components carry useful predictive information.

This study addresses these issues while keeping the framework of \citet{evkaya2021} as its starting point. It asks three related questions. First, does accounting for snow accumulation and melt change a drought index and its multiscale structure? Second, does multiscale information improve drought forecasts when it is constructed without any future information? Third, how much of any such improvement comes from the multiscale structure itself rather than from the longer history that the wavelet inputs summarize? To answer them, the SPI and the SMRI are computed for three stations in Ohio (Youngstown, Cleveland, and Columbus) over a common period, and their 3- and 6-month versions are used for forecasting. Each index is decomposed with the Haar maximal overlap discrete wavelet transform \citep[MODWT;][]{percival2000}, and the SPI and SMRI are compared component by component and month by month. In the forecasting experiment, wavelet components are ranked with Kendall's $\tau$ on the training period only, and the number of components is chosen by forecasting performance on a validation period rather than by a significance test. At every forecast origin, the wavelet inputs of the NARX model are recomputed from the history available before the target month only, and the models are re-estimated with an expanding window and evaluated one month ahead on a test period that is not used in any modeling decision. NAR and NARX are compared over 100 paired training repetitions, so that the added value of the multiscale information is judged against the variability caused by random network initialization. The forecasts are also compared with persistence, climatology, and linear autoregressive benchmarks, including an autoregression with a lag order chosen on the training period and the same autoregression with wavelet inputs, and differences in accuracy are tested with the Diebold--Mariano test.

This study makes three contributions. To our knowledge, it is the first comparison of the SPI and the SMRI in a wavelet--neural-network forecasting framework, in a region where snow is seasonal rather than dominant. It shows, scale by scale and month by month, where and when a snow-aware representation changes a drought index, and it separates the effect of the snow-adjusted input from that of the fitted distribution and of the snow-model parameters. Finally, it re-evaluates the wavelet--NAR/NARX framework with a leakage-free design, paired repetitions, and linear benchmarks of increasing memory, and shows that the forecast gain of the wavelet inputs is explained largely by the memory of the accumulated index.

The rest of the paper is organized as follows. Section~\ref{sec:data} describes the stations and data. Section~\ref{sec:methods} presents the methods: the drought indices and their comparison, the wavelet decomposition, the selection of wavelet components, the leakage-free forecasting design, and the evaluation of the forecasts. Section~\ref{sec:results} presents the results, Section~\ref{sec:discussion} discusses them, and Section~\ref{sec:conclusion} concludes.

\section{Study Area and Data}
\label{sec:data}

\subsection{Stations}
\label{sec:stations}

The study uses three weather stations in Ohio, United States: Youngstown--Warren Regional Airport (GHCN-Daily ID USW00014852) in northeastern Ohio, Cleveland Hopkins International Airport (USW00014820) near the southern shore of Lake Erie, and John Glenn Columbus International Airport (USW00014821) in central Ohio. The analysis was first developed for Youngstown and then applied to Cleveland and Columbus with the same code and settings. The three stations receive almost the same annual precipitation (about 1000\,mm) but differ in winter temperature and therefore in the role of snow (Table~\ref{tab:stations}). Youngstown is the coldest station, with the largest share of precipitation falling on days below 0\,$^\circ$C and the longest simulated snow cover; Columbus is the warmest, with about half the share of snow and the shortest snow cover; Cleveland lies in between. The stations therefore span a gradient of snow influence within a region where snow is seasonal rather than dominant.

\begin{table}[!htbp]
\centering
\caption{Climate and snow characteristics of the stations, 1953--2026. Snow share: percentage of precipitation falling on days with mean temperature below $T_T = 0\,^\circ$C. Maximum storage and snow days: mean annual maximum of the simulated snow water equivalent and mean number of days per year with at least 1\,mm, from the snow model of Section~\ref{sec:smri} (snow years July--June).}
\label{tab:stations}
\footnotesize
\begin{tabular}{lrrrrr}
\toprule
Station & Annual $P$ (mm) & DJF $T$ ($^\circ$C) & Snow share (\%) & Max.\ storage (mm) & Snow days \\
\midrule
Youngstown & 998 & $-2.2$ & 10.0 & 41 & 69 \\
Cleveland & 983 & $-1.4$ & 9.0 & 37 & 60 \\
Columbus & 1001 & $-0.5$ & 5.4 & 27 & 40 \\
\bottomrule
\end{tabular}
\end{table}


\subsection{Data and Preprocessing}
\label{sec:preprocessing}

Daily precipitation ($P$, mm) and daily maximum and minimum air temperature (TMAX, TMIN, $^\circ$C) were obtained from the Global Historical Climatology Network-Daily (GHCN-Daily) database \citep{menne2012} through the NOAA Climate Data Online service. Precipitation is the only input of the SPI and, together with temperature, the input of the snow model used for the SMRI. Temperature determines whether precipitation falls as snow or rain and when snow melts, and it is used as the daily mean $T=(\mathrm{TMAX}+\mathrm{TMIN})/2$. The downloaded files are kept unchanged; all later steps (choice of the study period, checks for missing values and absent days, filling of temperature gaps, and aggregation) are carried out by scripts that read these files, so that every result can be reproduced from the original data \citep{sandve2013,wilson2017}. The files and scripts are provided with the paper.


All three stations are analyzed over the common period from 1 January 1953 to 31 August 2026 (884 complete months; the incomplete September 2026 is excluded). The start is set by Youngstown, whose record has absent calendar days in October 1944, the whole of 1947, and October 1952; the Columbus and Cleveland records have no absent days from earlier dates (January 1948 and May 1938, respectively). A common period is used because both indices are standardized with distributions fitted to the record itself, so that index values, as well as the training, validation, and test periods (Section~\ref{sec:forecast}), refer to the same months at every station. The period is about 74 years long, well above the minimum of 30 years recommended for the SPI \citep{svoboda2012}.

Within the common period, the records of all three stations were checked for missing values and for calendar days absent from the file. No station has missing precipitation or absent days. The temperature records of Columbus and Cleveland are complete; at Youngstown, a single value is missing (TMAX on 16 April 1997). Missing precipitation would not have been filled, because daily precipitation has weak day-to-day dependence and is zero on many days, so that a value interpolated in time would not be a plausible estimate; the scripts stop if a missing precipitation value or an absent day is found. Temperature varies smoothly from day to day, and the single missing TMAX value is filled by linear interpolation in time. The checks, including the records before 1953, are reported in Supplementary Section~A.

Daily precipitation is summed to monthly totals for the SPI. For the SMRI, the snow model is run on the \emph{daily} series, and the resulting daily snowmelt-plus-rain is then summed to monthly totals (Section~\ref{sec:smri}). The snow processes are thus simulated at the daily time step at which they occur, before any monthly aggregation.

\section{Methods}
\label{sec:methods}

The analysis proceeds in the following steps. The SPI, the SMRI, and auxiliary versions with the other distribution are computed (Sections~\ref{sec:spi}--\ref{sec:spip3}), and the SPI and SMRI are compared (Section~\ref{sec:repcomp}). Each index is decomposed with the Haar MODWT (Section~\ref{sec:modwt}), and the wavelet components are screened and selected (Section~\ref{sec:selection}). NAR and NARX forecasts are then produced with a leakage-free design, together with linear benchmarks (Section~\ref{sec:forecast}), and evaluated (Section~\ref{sec:eval}). All computations are carried out in R version 4.6.1 \citep{rcore}.


\subsection{Standardized Precipitation Index}
\label{sec:spi}

Two drought indices are compared: the SPI, the index used by \citet{evkaya2021}, and the SMRI \citep{staudinger2014} (Section~\ref{sec:smri}). Both are computed with the same standardization procedure (accumulation over $k$ months, a distribution fitted separately for each calendar month, and transformation to a standard normal variable) and over the same study period. They differ in the input variable and in the fitted distribution: the SPI uses precipitation as it falls and, as in \citet{evkaya2021}, the Gamma distribution, whereas the SMRI uses the water that actually reaches the ground after snow has been stored and melted and, as in \citet{staudinger2014}, the Pearson type~III distribution. To separate these two differences, each index is also computed with the other distribution (Section~\ref{sec:spip3}); these auxiliary versions are used only as a check and not for forecasting. Both indices are computed at accumulation periods of $k = 3$ and $k = 6$ months, which are used throughout the study. Where a specific accumulation period is meant, the index is written SPI-3, SMRI-6, and so on; $\mathrm{SPI}_k$ and $\mathrm{SMRI}_k$ denote the index at a generic accumulation period $k$. The 9- and 12-month indices are also computed, but they are used only in the descriptive comparison (Section~\ref{sec:repcomp}).

The SPI \citep{mckee1993} is computed in three steps. First, monthly precipitation is accumulated over a moving window of $k$ months. Second, a probability distribution is fitted to the accumulated totals separately for each calendar month. Third, the fitted cumulative probabilities are mapped to the standard normal distribution. Following \citet{evkaya2021}, the two-parameter Gamma distribution
\begin{equation}
g(x)=\frac{1}{\beta^{\alpha}\Gamma(\alpha)}\,x^{\alpha-1}e^{-x/\beta},\qquad x>0,
\end{equation}
is used, with shape $\alpha>0$ and scale $\beta>0$. With $G(x)$ the fitted cumulative distribution function and $q$ the probability of a zero total,
\begin{equation}
\mathrm{SPI}=\Phi^{-1}\big(\tilde G(x)\big),\qquad \tilde G(x)=q+(1-q)\,G(x),
\end{equation}
where $\Phi$ is the standard normal distribution function. The index is computed with the \texttt{spi()} function of the R package \textbf{SPEI} \citep{spei}, using the whole study period as the reference period. The distribution parameters are estimated with unbiased probability-weighted moments, the default of \texttt{spi()}. Negative values indicate conditions drier than normal for the calendar month, and positive values indicate wetter conditions. Following the classification of \citet{mckee1993}, values below $-1$ indicate moderate or more severe drought.

\subsection{Standardized Snow Melt and Rain Index}
\label{sec:smri}

The SMRI applies the SPI procedure to the daily sum of snowmelt and rain (MR) instead of precipitation. In winter, precipitation can be stored in the snowpack and released weeks later as melt. MR can therefore differ markedly from precipitation in \emph{when} water reaches the ground, even when the total amounts are similar. Below, the \emph{snow-adjusted input} is MR, and a \emph{snow-aware} index is an index computed from it.

MR is simulated with the degree-day snow model of \citet[Appendix~A]{staudinger2014}, implemented in R directly from the published algorithm. Let $A_t$ be the snow storage (snow water equivalent), $W_t$ the liquid water held in the snowpack, and $Q_t$ the water released to the ground on day $t$. On a cold day ($T_t < T_T$), precipitation is added to the snowpack and part of the liquid water refreezes:
\begin{equation}
\begin{gathered}
F_t=\min\{W_{t-1},\;C_{FR}\,C_M\,(T_T-T_t)\},\\
A_t=A_{t-1}+P_t+F_t,\qquad W_t=W_{t-1}-F_t,\qquad Q_t=0.
\end{gathered}
\end{equation}
On a warm day ($T_t \ge T_T$), snow melts and joins the rain as liquid water:
\begin{equation}
M_t=\min\{A_{t-1},\;C_M\,(T_t-T_T)\},\quad A_t=A_{t-1}-M_t,\quad W_t=W_{t-1}+P_t+M_t.
\end{equation}
The snowpack can hold liquid water up to a fraction $C_{WH}$ of its snow storage, and any excess leaves the snowpack:
\begin{equation}
Q_t=\max\{0,\;W_t-C_{WH}A_t\},\qquad W_t\leftarrow\min\{W_t,\;C_{WH}A_t\}.
\end{equation}
Here $T_T$ is the threshold temperature, $C_M$ the degree-day factor, $C_{FR}$ the refreezing coefficient, and $C_{WH}$ the water-holding capacity of the snowpack. They are fixed at $T_T = 0\,^\circ$C, $C_M = 3$\,mm\,$^\circ$C$^{-1}$\,d$^{-1}$, $C_{FR} = 0.05$, and $C_{WH} = 0.1$, which are typical values for degree-day snow models; the sensitivity of the results to these values is examined in Section~\ref{sec:repcomp}. The model starts with an empty snowpack ($A_0 = W_0 = 0$). The code was verified on hand-calculated examples, including a check that precipitation equals released water plus storage over the whole run.

Daily $\mathrm{MR}_t = Q_t$ is summed to monthly totals, accumulated over $k$ months, and standardized with the same \texttt{spi()} routine, using the Pearson type~III distribution as in \citet{staudinger2014}. For both indices, the first $k-1$ values cannot be computed and are removed. Because all stations share the study period, this leaves at every station $n = 882$ values for $\SPI{3}$ and $\SMRI{3}$ (March 1953--August 2026) and $n = 879$ for $\SPI{6}$ and $\SMRI{6}$ (June 1953--August 2026).

A few of the most extreme months give very large index values. In addition, the fitted Pearson type~III distribution has a lower bound, and when an accumulated total lies below it the index is $-\infty$; among the forecast series, this happens once for $\SMRI{6}$ at Columbus, and twice for $\SMRI{3}$ and three times for $\SMRI{6}$ at Cleveland. Following \citet[Section~2.5]{stagge2015}, all index values (SPI, SMRI, and the auxiliary versions of Section~\ref{sec:spip3}) are therefore bounded to the interval $[-3, 3]$, which corresponds to a probability of about 0.14\% in each tail, or roughly one event in 740 years for a given calendar month. The same rule is applied to every index, accumulation period, and station. Over the four forecast series ($\SPI{3}$, $\SMRI{3}$, $\SPI{6}$, $\SMRI{6}$), 8 of the 3522 values per station are bounded at Youngstown, 13 at Columbus, and 18 at Cleveland; the bounded months are listed in Supplementary Table~S1.

\subsection{Distribution Check}
\label{sec:spip3}

The SPI and the SMRI differ in two respects at once: the input variable (precipitation vs.\ snowmelt plus rain) and the fitted distribution (Gamma vs.\ Pearson type~III). A difference between the two indices could therefore reflect the choice of distribution rather than the effect of snow. To separate the two effects, each index is also computed with the other distribution, using the same \texttt{spi()} routine and settings: the SPI with the Pearson type~III distribution ($\mathrm{SPI}^{\mathrm{P3}}_k$) and the SMRI with the Gamma distribution ($\mathrm{SMRI}^{\mathrm{G}}_k$). Together with the SPI and the SMRI, this gives all four combinations of input and distribution. Comparing indices with the same distribution ($\mathrm{SMRI}_k$ vs.\ $\mathrm{SPI}^{\mathrm{P3}}_k$, and $\mathrm{SMRI}^{\mathrm{G}}_k$ vs.\ $\mathrm{SPI}_k$) isolates the effect of the snow-adjusted input, and comparing indices with the same input ($\mathrm{SPI}^{\mathrm{P3}}_k$ vs.\ $\mathrm{SPI}_k$, and $\mathrm{SMRI}_k$ vs.\ $\mathrm{SMRI}^{\mathrm{G}}_k$) isolates the effect of the distribution, for precipitation and for snowmelt plus rain separately (Section~\ref{sec:repcomp}). The auxiliary indices are used only for this check; they are not decomposed with wavelets or used in the forecasting experiment.

\subsection{Comparison of SPI and SMRI}
\label{sec:repcomp}

The two representations are compared at the same accumulation period ($\mathrm{SPI}_k$ vs.\ $\mathrm{SMRI}_k$, $k = 3, 6, 9, 12$). At the input level, monthly precipitation and monthly MR are compared with Pearson's correlation, to check whether the snow adjustment mainly redistributes water in time or changes the amount. The overall agreement between the indices is measured with Pearson's $r$ and Kendall's $\tau$, and the mean of the difference $\delta_t = \mathrm{SMRI}_k(t) - \mathrm{SPI}_k(t)$ shows whether the SMRI is shifted systematically.

The seasonal pattern of the differences is summarized by the mean absolute difference for each calendar month $m$,
\begin{equation}
\overline{|\delta|}_m=\frac{1}{n_m}\sum_{t:\,\mathrm{month}(t)=m}\big|\mathrm{SMRI}_k(t)-\mathrm{SPI}_k(t)\big| ,
\end{equation}
where $n_m$ is the number of years in which month $m$ is observed. Because a mean can be dominated by a few extreme months, the same summary is also computed with the median instead of the mean. Comparing $\overline{|\delta|}_m$ across $k$ shows whether a longer accumulation window weakens or redistributes the seasonal differences.

Two further checks examine whether the differences can be attributed to the snow-adjusted input. The first concerns the distribution. With the auxiliary indices of Section~\ref{sec:spip3}, four further monthly differences are computed with the same formula: $\mathrm{SMRI}_k - \mathrm{SPI}^{\mathrm{P3}}_k$ and $\mathrm{SMRI}^{\mathrm{G}}_k - \mathrm{SPI}_k$, in which only the input differs, and $\mathrm{SPI}^{\mathrm{P3}}_k - \mathrm{SPI}_k$ and $\mathrm{SMRI}_k - \mathrm{SMRI}^{\mathrm{G}}_k$, in which only the distribution differs. The correlation between indices with the same input and different distributions is also reported. If the first two retain the seasonal pattern of $\delta_t$ and the last two are small in every month, the seasonal differences can be attributed to the snow-adjusted input rather than to the choice of distribution. The second check concerns the snow-model parameters. For $k = 3$ and $6$, the SMRI is recomputed with 200 parameter sets drawn independently and uniformly from $T_T \in [-2.5, 2.5]\,^\circ$C, $C_M \in [1, 6]$\,mm\,$^\circ$C$^{-1}$\,d$^{-1}$, and $C_{FR} \in [0, 0.1]$, with $C_{WH}$ fixed at 0.1, and bounded to $[-3, 3]$ in the same way as the other indices. For each set, $\overline{|\delta|}_m$ is computed against the SPI, and the 5th, 50th, and 95th percentiles over the 200 sets are compared with the values obtained with the fixed parameters.

These comparisons are descriptive. They show where and when the representations differ, but they do not prove that snow processes cause the differences. In the following sections, a difference is called attributable to the snow-adjusted input when it is reproduced by the pairs of indices that differ only in the input.

\subsection{Wavelet Decomposition}
\label{sec:modwt}

As in \citet{evkaya2021}, a wavelet decomposition is used as a preprocessing step that separates each index into components describing its variation at different time scales. The Haar wavelet is kept for continuity with \citet{evkaya2021}, but the decimated discrete wavelet transform (DWT) used there is replaced by the maximal overlap discrete wavelet transform \citep[MODWT;][]{percival2000}. Let $X_t$, $t = 1,\dots,N$, be an index series, and let $\tilde h_{j,l}$, $l = 0,\dots,K_j-1$, be the level-$j$ MODWT wavelet filter. The level-$j$ wavelet coefficients are
\begin{equation}
\widetilde W_{j,t}=\sum_{l=0}^{K_j-1}\tilde h_{j,l}\,X_{t-l}.
\end{equation}
For the Haar wavelet, $K_j = 2^j$, and the coefficient is, up to sign and scale, the difference between the means of two adjacent blocks of $2^{j-1}$ observations. The coefficients are converted back to the time domain to give a multiresolution analysis (MRA):
\begin{equation}
X_t=\sum_{j=1}^{6}L_{j,t}+S_{6,t},\qquad
L_{j,t}=\sum_{l=0}^{K_j-1}\tilde h_{j,l}\,\widetilde W_{j,t+l},
\label{eq:mra}
\end{equation}
where $L_{j,t}$ is the detail component at level $j$, referred to below as \emph{wavelet level} $L_j$ (written $L_j(t)$ when its value in month $t$ is meant), and $S_{6,t}$ is the smooth component. Six levels are used, as in \citet{evkaya2021}; $L_1$ represents the finest and $L_6$ the coarsest time scale. Exact period interpretations are avoided because the Haar filter separates frequencies only roughly.

The MRA is computed with the \texttt{mra()} function of the \textbf{waveslim} package \citep{waveslim}, using the Haar filter, $J = 6$, and reflection boundary conditions ($L_j$ corresponds to the component $D_j$ in the output of \texttt{mra()}). For every series, it is checked numerically that the components add up to the original series as in \eqref{eq:mra}. The MODWT is preferred to the DWT for three reasons: it can be applied to series of any length, so the 882- and 879-month series do not have to be truncated to a power of two; every level keeps $N$ values aligned in time with the index; and the result does not depend on where the series starts. Together, these properties allow the SPI and SMRI to be compared level by level and month by month.

To examine whether the snow-adjusted input changes the multiscale information, the full-record decompositions of the SPI and SMRI are compared level by level at the same accumulation period ($L_j^{\mathrm{SPI}}$ vs.\ $L_j^{\mathrm{SMRI}}$, $j = 1,\dots,6$, and $S_6$) for $k = 3$ and $6$. Three measures are used: Pearson's correlation, Kendall's $\tau$, and the mean absolute difference. The mean absolute difference between corresponding levels is also computed for each calendar month, to show whether representation differences at a given scale are seasonal. Finally, the relative energy of each component, $E_C = \sum_t C_t^2 / \sum_{C'}\sum_t C_t'^2$ with $C, C' \in \{L_1,\dots,L_6,S_6\}$, describes how the variability of each index is distributed across scales. Because the MODWT is a redundant transform, $E_C$ is used as a descriptive measure and not as an exact share of the variance.

Equation~\eqref{eq:mra} also shows that $L_{j,t}$ depends on coefficients at later times, $\widetilde W_{j,t+l}$, and therefore on observations after time $t$. Decompositions of the full record are therefore used \emph{only} for description (this section and the association analysis in Section~\ref{sec:selection}). The inputs of the forecasting models are built from past data only (Section~\ref{sec:forecast}).

\subsection{Selection of Wavelet Components}
\label{sec:selection}

\citet{evkaya2021} selected wavelet series by testing the correlation between each series and the original index with Kendall's $\tau$. For two series $(x_t, y_t)$, $t = 1,\dots,n$, Kendall's $\tau$ \citep{kendall1938} is
\begin{equation}
\tau=\frac{n_c-n_d}{n(n-1)/2},
\end{equation}
where $n_c$ and $n_d$ are the numbers of concordant and discordant pairs. For continuity with \citet{evkaya2021}, this rule is applied first to the full-record decomposition of each index: a level is retained if its Kendall correlation with the index is significant at the 5\% level (large-sample normal approximation). Two features of the present data limit the usefulness of this rule. First, with about 880 monthly values, even weak correlations become significant. Second, the series are serially dependent, which the usual test ignores. The uncertainty of $\tau$ is therefore also assessed with a paired moving-block bootstrap \citep{kunsch1989}: blocks of consecutive months, starting at randomly chosen positions, are drawn with replacement and concatenated until the original length is reached, keeping the index and the wavelet level together. Percentile 95\% confidence intervals are computed from $B = 1000$ resamples for every index and every level. Because the result can depend on the block length, block lengths of 6, 12, and 24 months are compared. If the significance rule retains all, or almost all, levels, it cannot tell which levels are useful, and $\tau$ is then used only to rank the levels.

The number of wavelet levels used for forecasting is therefore chosen by predictive performance rather than by a significance test or a fixed cutoff on $|\tau|$. First, the training part of each index series (Section~\ref{sec:forecast}) is decomposed with the MODWT, and $L_1,\dots,L_6$ are ranked by $|\tau|$ with the training-period index. Second, six nested candidate sets are formed from this ranking: Top-1 contains the highest-ranked level, Top-2 the two highest, and so on up to Top-6, which contains all six levels. Third, each set is used as the input of a NARX model (and of the linear benchmark with wavelet inputs), and the set with the best validation performance is selected (Section~\ref{sec:forecast}). The association analysis thus shows which scales are associated with the index, whereas the prediction-driven selection shows which scales actually improve the forecast.

\subsection{Leakage-Free Forecasting Design}
\label{sec:forecast}

As in \citet{evkaya2021}, two dynamic neural-network models are used, and the same design is applied separately to each forecast series ($\SPI{3}$, $\SMRI{3}$, $\SPI{6}$, $\SMRI{6}$). The NAR model predicts the index $I(t)$ from its own past values. With delay $d = 2$, as in \citet{evkaya2021},
\begin{equation}
I(t)=f\big(I(t-1),\,I(t-2)\big)+\varepsilon_t ,
\label{eq:nar}
\end{equation}
where $f(\cdot)$ is a nonlinear function estimated by the network and $\varepsilon_t$ is the error. The NAR model is the baseline and shows how well each index can be predicted from its own recent past alone. The NARX model adds the past values of the selected wavelet levels as exogenous inputs:
\begin{equation}
I(t)=f\Big(I(t-1),\,I(t-2),\;\big\{L^{(t-1)}_{j}(t-1),\,L^{(t-1)}_{j}(t-2)\big\}_{j\in\mathcal S}\Big)+\varepsilon_t ,
\label{eq:narx}
\end{equation}
where $\mathcal S$ is a candidate set (Top-1 to Top-6), so that the network has $2 + 2|\mathcal S|$ inputs. The superscript $(t-1)$ indicates that the levels are computed from data up to $t-1$ only. The same delay is used for the index and the wavelet levels, so the difference between NAR and NARX measures what the multiscale information adds.

For an input vector $\mathbf x$, $f(\cdot)$ is a feed-forward network with one hidden layer of $H$ logistic neurons and a linear output neuron:
\begin{equation}
f(\mathbf x)=w_0+\sum_{h=1}^{H}w_h\,\sigma\!\left(v_{0h}+\mathbf v_h^{\top}\mathbf x\right),\qquad
\sigma(u)=\frac{1}{1+e^{-u}},
\end{equation}
with $H = 4$, $5$, or $6$, as in \citet{evkaya2021}. The networks are fitted with the \textbf{nnet} package (version 7.3-20) \citep{venables2002}. The weights minimize the sum of squared errors and are found with the BFGS quasi-Newton method, for at most 1000 iterations, without weight decay, and from random starting weights. This differs from the Levenberg--Marquardt algorithm used in MATLAB by \citet{evkaya2021}. Before each fit, the inputs and the target are standardized with the means and standard deviations of the current training rows; the same constants are applied to the inputs of the target month, and the forecast is transformed back to the index scale.

Each index series is split chronologically into a training period (first 60\%), a validation period (next 20\%), and a test period (last 20\%; Table~\ref{tab:split}). The training period is used to rank the wavelet levels (Section~\ref{sec:selection}) and for the first estimation of the network weights. The validation period is used to choose the number of hidden neurons for NAR, and the Top-$m$ set together with the number of hidden neurons for NARX. The test period is used only once, after all choices have been fixed, for the final out-of-sample comparison. Test observations are never used for ranking, selection, or tuning.

\begin{table}[!htbp]
\centering
\caption{Chronological 60/20/20 partition of the index series (number of months and period; the same for all three stations).}
\label{tab:split}
\footnotesize
\setlength{\tabcolsep}{3.5pt}
\begin{tabular}{lrlll}
\toprule
Series & $n$ & Training & Validation & Test \\
\midrule
$\SPI{3}$, $\SMRI{3}$ & 882 & 529 (Mar 1953--Mar 1997) & 176 (Apr 1997--Nov 2011) & 177 (Dec 2011--Aug 2026)\\
$\SPI{6}$, $\SMRI{6}$ & 879 & 527 (Jun 1953--Apr 1997) & 176 (May 1997--Dec 2011) & 176 (Jan 2012--Aug 2026)\\
\bottomrule
\end{tabular}
\end{table}

The wavelet inputs are constructed from past information only. For every target month $t$, the MRA \eqref{eq:mra} is computed from the history $\{I(1),\dots,I(t-1)\}$, and the last two values of each selected level, $L^{(t-1)}_j(t-1)$ and $L^{(t-1)}_j(t-2)$, are taken as inputs. The inputs used to forecast $I(t)$ therefore never contain $I(t)$ or any later observation. Because the MRA value at a given time also depends on coefficients at later times \eqref{eq:mra}, the last values of a decomposition of $\{I(1),\dots,I(t-1)\}$ are obtained with the reflection boundary rule, which extends the history by its mirror image rather than by future observations. These end-of-history values differ from the values that the same level takes in a decomposition of the full record. This is intended: they are the information that would actually be available at the forecast origin. Every row of the NARX training data is constructed in the same way, so the inputs behave the same in training as in a real forecast. At least 128 past months are required before a decomposition is computed, so the NARX rows start from the 129th month of each series.

In the validation and test periods, forecasts are made one month ahead with an expanding window. For each target month, the network is re-estimated on all rows available before that month and then forecasts $I(t)$; once $I(t)$ is observed, it is added to the data for the next month. Forecasts use the observed past values rather than the model's own earlier forecasts, which corresponds to the series--parallel arrangement described by \citet{evkaya2021}, and the input and target standardization is recomputed at each step. NAR and NARX are evaluated on exactly the same target months, but they are not estimated from the same rows. The NAR model uses all target months from the third month of the series onward, whereas the NARX model can use target months only from the 129th month onward, because each of its rows requires 128 months of history for the decomposition. At the first validation month of $\SPI{3}$, for example, the NAR model is estimated from 527 rows and the NARX model from 401. The NARX model therefore has no advantage in the amount of estimation data.

Because the fitted network depends on the random starting weights, every specification (index $\times$ model $\times$ Top-$m$ set $\times$ $H$) is evaluated over $R = 100$ independent repetitions, in line with \citet{evkaya2021}. In the final test, NAR and NARX use the same random seed in each repetition $i$, so that the two models can be compared repetition by repetition. A small number of starting weights can lead to failed fits with very large errors, which can pull the mean over repetitions far from the typical result. Performance across repetitions is therefore summarized mainly by the median, with the interquartile range (IQR) and the standard deviation describing the spread. The single best repetition is never used, because it reflects a fortunate set of starting weights rather than the model.

Model selection uses the median validation RMSE over the 100 repetitions (Section~\ref{sec:eval}) as the sole criterion: for NAR, it chooses $H$; for NARX, it chooses the Top-$m$ set and $H$ jointly. The median validation MAE and the IQR and standard deviation of the RMSE are reported as secondary information. This rule was fixed before the NARX experiments and is applied in the same way to every index and station. The indices themselves are standardized with distributions fitted to the whole study period (Sections~\ref{sec:spi} and \ref{sec:smri}), as is usual for the SPI. This affects the target series in the same way for all models; the leakage-free design concerns the forecast inputs and all modeling decisions.

The NAR and NARX models see only the last two values of the index, whereas the wavelet inputs summarize up to 128 months of history. To separate the effect of the multiscale structure from that of a longer memory, two linear benchmarks are added. The first is an autoregression with an intercept, AR($p$), whose order $p \in \{1,\dots,12\}$ is chosen by the Akaike information criterion \citep{akaike1974} on the training period (on a common estimation sample) and then kept fixed. The second, AR($p$)+W, adds to AR($p$) the two lagged values of wavelet levels from the same causal decompositions as in NARX; the set of levels is chosen among ``none'' and the Top-1 to Top-6 sets of Section~\ref{sec:selection} by the validation RMSE. Both are estimated by least squares with the same expanding window and on the same rows as NARX, so that they differ only in the wavelet inputs; coefficients of inputs that are exact linear combinations of other inputs are set to zero. They are deterministic and need no repetitions. Because the lag order and the wavelet levels are chosen with the same rules as for the networks and the test period is used only once, the benchmarks are subject to the same leakage-free design.

The two-lag networks and the autoregressions differ in memory, so the 3-month series are also modeled with networks that have more index lags. The number of lags $d$ of the NAR model is chosen among 2, 6, and 12 by the median validation RMSE over 100 repetitions, with $H = 4$. The NARX model uses the same $d$ for the index and the wavelet levels selected for the two-lag model, each of which keeps its two lags. The 6-month series, with no wavelet gain at two lags, are not repeated.

\subsection{Forecast Evaluation}
\label{sec:eval}

Let $I(t)$ be the observed index and $\hat I_i(t)$ its forecast in repetition $i$, over the $n_P$ target months of an evaluation period. Forecast accuracy is measured by the root mean squared error and the mean absolute error,
\begin{equation}
\mathrm{RMSE}_i=\sqrt{\frac{1}{n_P}\sum_{t}\big(I(t)-\hat I_i(t)\big)^2},\qquad
\mathrm{MAE}_i=\frac{1}{n_P}\sum_{t}\big|I(t)-\hat I_i(t)\big|.
\end{equation}
Both are expressed in index units, that is, in standard deviations of the index. RMSE gives more weight to large errors, such as those in extreme dry or wet months, whereas MAE weights all errors equally. For comparison with \citet{evkaya2021}, the normalized mean absolute error and the similarity measure used there are also reported:
\begin{equation}
\mathrm{NMAE}=\frac{1}{n_P}\sum_{t}\frac{|I(t)-\hat I(t)|}{I_{\max}-I_{\min}},\qquad
\mathrm{SIM}=\frac{1}{n_P}\sum_{t}\frac{1}{1+\dfrac{|I(t)-\hat I(t)|}{\hat I_{\max}-\hat I_{\min}}},
\end{equation}
where $I_{\max}$ and $I_{\min}$ are the largest and smallest observed values and $\hat I_{\max}$ and $\hat I_{\min}$ the largest and smallest forecasts in the evaluation period. Because SIM divides by the range of the forecasts, it is not informative for nearly constant forecasts such as climatology.

The value added by the wavelet levels is measured within each index, by comparing NAR and NARX for the same index on the same test months. For a loss $\mathcal L \in \{\mathrm{RMSE}, \mathrm{MAE}\}$ and repetition $i$,
\begin{equation}
\Delta\mathcal L_i=\mathcal L^{\mathrm{NAR}}_i-\mathcal L^{\mathrm{NARX}}_i ,
\end{equation}
where positive values indicate that the multiscale information improved the forecast. Three summaries are reported: the median and IQR of $\Delta\mathcal L_i$; the relative gain, $100 \times \mathrm{median}(\Delta\mathcal L_i) / \mathrm{median}(\mathcal L^{\mathrm{NAR}}_i)$; and the proportion of repetitions in which NARX is better, $\pi = R^{-1}\sum_{i=1}^{R}\mathbf 1(\Delta\mathcal L_i>0)$, which shows whether the gain holds across starting weights. The number of repetitions with an RMSE above 1 is also reported for each model, to identify failed fits.

In addition to the individual repetitions, an ensemble forecast is formed for each model by taking, for every test month, the median of the 100 forecasts. The ensemble forecasts are compared with five reference forecasts computed for the same test months: persistence, $\hat I(t) = I(t-1)$; climatology, the mean of all earlier values of the index; a linear autoregressive model of order 2, AR(2), with an intercept, estimated by least squares with the same expanding window; and the AR($p$) and AR($p$)+W benchmarks of Section~\ref{sec:forecast}. The reference forecasts show whether the networks improve on simple benchmarks.

Differences in accuracy between two forecasts of the same months are tested with the Diebold--Mariano test \citep{diebold1995} with the small-sample correction of \citet{harvey1997}, for one-step-ahead forecasts and with the corrected statistic compared with Student's $t$ distribution with $n_P - 1$ degrees of freedom. The test is applied with squared-error and absolute-error loss to the pairs NAR vs.\ NARX ensemble, persistence vs.\ NAR and NARX ensemble, AR(2) vs.\ NAR and NARX ensemble, AR($p$) vs.\ NAR and NARX ensemble, and AR($p$) vs.\ AR($p$)+W. In addition, the test (squared-error loss) is applied to NAR and NARX in every repetition, and the proportion of repetitions in which NARX is better and the difference is significant at the 5\% level is reported.

Because accurate forecasts of dry conditions matter most for early warning, the forecasts are also evaluated as drought warnings. A month is a drought month when the index is below $-1$ (Section~\ref{sec:spi}). Comparing observed and forecast drought months gives the numbers of hits, misses, and false alarms, from which the probability of detection, $\mathrm{POD} = \text{hits}/(\text{hits}+\text{misses})$, and the false alarm ratio, $\mathrm{FAR} = \text{false alarms}/(\text{hits}+\text{false alarms})$, are computed for the ensemble and reference forecasts. Because the test period contains only 4--16 drought months per series, these scores are reported descriptively.

Finally, because the SPI and SMRI differ mainly in winter and spring, the test months are grouped by meteorological season (December--February, March--May, June--August, September--November). For each season, the median RMSE of NAR and NARX over the repetitions, the median $\Delta\mathrm{RMSE}_i$, the proportion of repetitions in which NARX is better, and the RMSE of the persistence and AR(2) forecasts are computed.

Each index is a different target series, so the absolute error levels of the SPI and SMRI are not compared directly. Instead, the effect of the drought representation is assessed through (i) which wavelet levels are selected and (ii) whether, and how consistently, $\Delta\mathcal L$ is positive for each index.

\section{Results}
\label{sec:results}

\subsection{Agreement Between SPI and SMRI}
\label{sec:res_agree}

Over the whole record, the SPI and SMRI are almost interchangeable. Monthly snowmelt plus rain is highly correlated with monthly precipitation at all three stations ($r = 0.958$ at Youngstown, 0.967 at Cleveland, and 0.984 at Columbus). The SPI and SMRI therefore agree closely overall (Supplementary Table~S2). At the 3-month accumulation period, Pearson's $r$ between the two indices is 0.974 at Youngstown, 0.980 at Cleveland, and 0.988 at Columbus, and Kendall's $\tau$ is 0.887, 0.898, and 0.929. Agreement increases with the accumulation period, to $r = 0.995$--0.998 and $\tau = 0.961$--0.973 at 12 months. The mean difference $\mathrm{SMRI}_k - \mathrm{SPI}_k$ is below 0.003 in absolute value for every station and accumulation period, so the SMRI is not shifted systematically relative to the SPI. The choice of distribution alone has little effect on the overall agreement: the correlation between indices with the same input and different distributions is 0.996--0.999 for both precipitation ($\mathrm{SPI}_k$ vs.\ $\mathrm{SPI}^{\mathrm{P3}}_k$) and snowmelt plus rain ($\mathrm{SMRI}_k$ vs.\ $\mathrm{SMRI}^{\mathrm{G}}_k$).

\subsection{Seasonal Differences Between SPI and SMRI}
\label{sec:res_season}

The high overall agreement hides a clear seasonal signal: at the 3-month accumulation period, the two indices differ mainly in winter and spring, and the difference is attributable to the snow-adjusted input (Fig.~\ref{fig:season}; the values are given in Supplementary Tables~S3--S5). At the 3-month accumulation period, the mean absolute difference between SMRI and SPI is largest from January to May, with maxima of 0.28 at Youngstown (January), 0.25 at Cleveland (February), and 0.17 at Columbus (January), and it does not exceed 0.11 from June to November. Averaged over the twelve calendar months, it is 0.133 at Youngstown, 0.124 at Cleveland, and 0.087 at Columbus. From December to May, the differences in which only the input changes ($\mathrm{SMRI}_k - \mathrm{SPI}^{\mathrm{P3}}_k$ and $\mathrm{SMRI}^{\mathrm{G}}_k - \mathrm{SPI}_k$) are as large as the total difference, whereas the differences in which only the distribution changes remain at or below 0.11 at Youngstown, 0.04 at Cleveland, and 0.09 at Columbus. The winter and spring differences at the 3-month accumulation period can therefore be attributed to the snow-adjusted input. From June to November, the input-only differences do not exceed 0.05, and the smaller differences that remain in these months, such as the August value of 0.11 at Youngstown, are attributable to the choice of distribution.

With longer accumulation periods, the differences become smaller on average (0.052 at Youngstown, 0.043 at Cleveland, and 0.044 at Columbus at 12 months) and spread into the following months, because the accumulation window then contains the snow season; at 6 months, for example, input-only differences reach 0.15 at Youngstown and 0.14 at Cleveland in July and August. At the same time, the relative role of the distribution increases. At Cleveland and Columbus, the distribution-only differences at 6 and 9 months are as large as the total difference in some winter and spring months (for example, 0.18 vs.\ 0.19 in December at Cleveland and 0.15 vs.\ 0.15 in March at Columbus at 6 months), while the input-only differences in these months are much smaller (0.05--0.07 at Cleveland in December and below 0.01 at Columbus in March). The median absolute differences show the same pattern (Supplementary Tables~S3--S5), so these months are not driven by a single extreme year. At 12 months, the input-only differences are below 0.03 from June to November at all stations, and the remaining differences in these months are of the same size as the distribution-only differences.

\begin{figure}[!htbp]
\centering
\includegraphics[width=\textwidth]{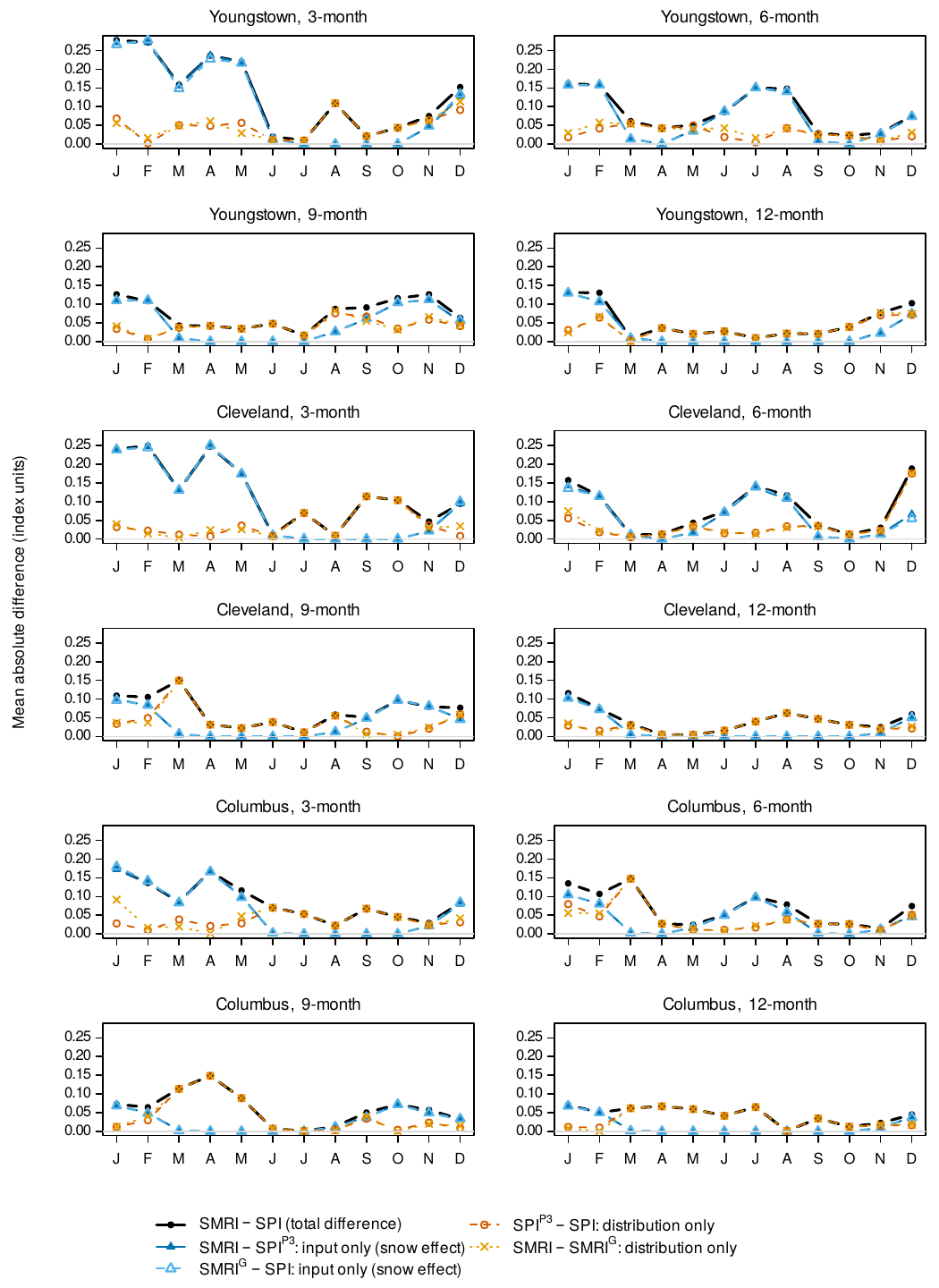}
\caption{Mean absolute difference between indices by calendar month, for each station and accumulation period (two rows per station: 3 and 6 months, 9 and 12 months). Black: SMRI vs.\ SPI. Blue: pairs that differ only in the input (snowmelt plus rain vs.\ precipitation). Orange: pairs that differ only in the distribution (Gamma vs.\ Pearson type~III). Where the blue lines follow the black line, the difference between SMRI and SPI is attributable to the snow-adjusted input; where the orange lines follow it, the difference is attributable to the choice of distribution. $\mathrm{SPI}^{\mathrm{P3}}$: SPI with the Pearson type~III distribution; $\mathrm{SMRI}^{\mathrm{G}}$: SMRI with the Gamma distribution.}
\label{fig:season}
\end{figure}

\begin{figure}[!htbp]
\centering
\includegraphics[width=\textwidth]{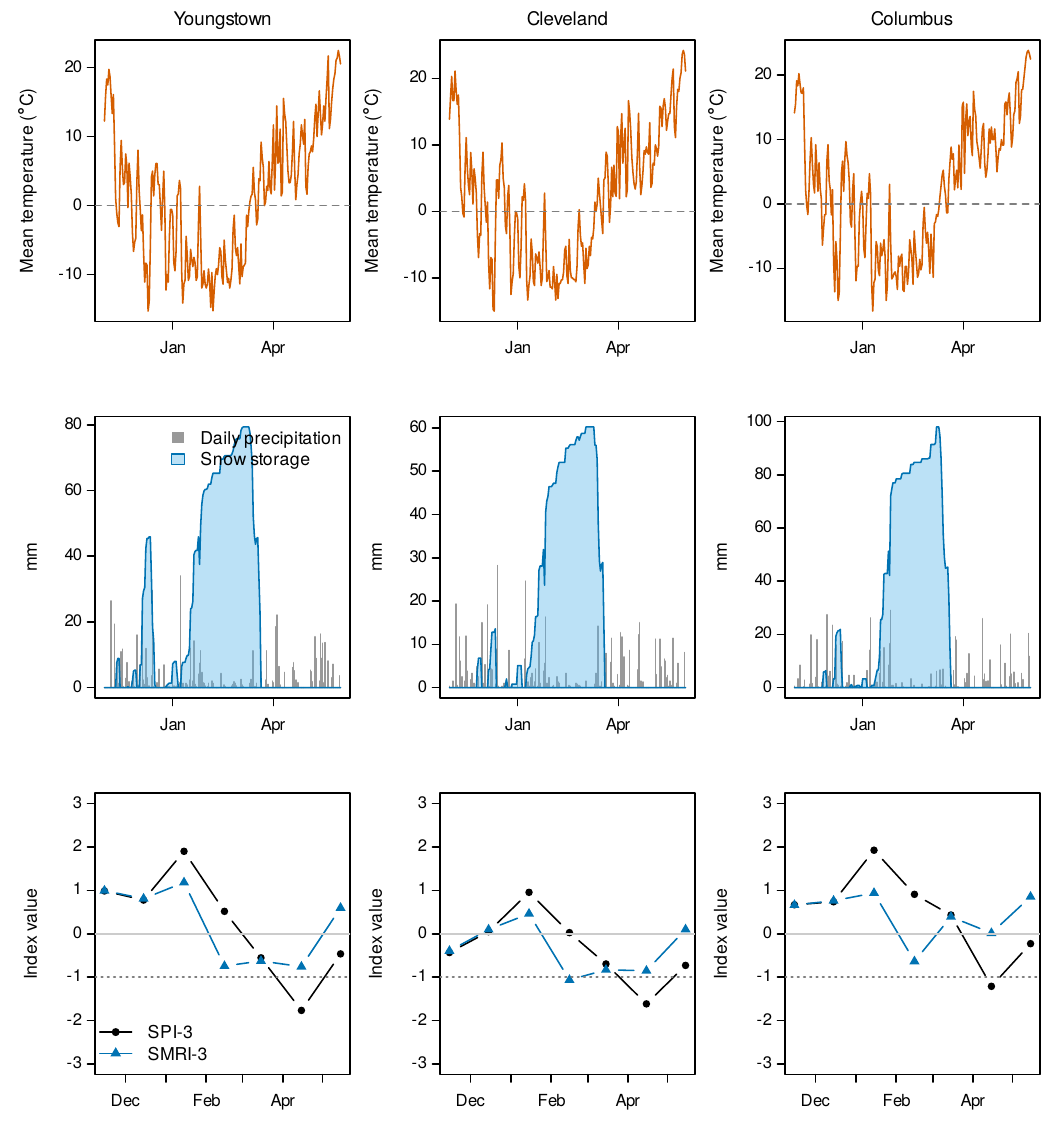}
\caption{Winter 1977--78 at the three stations. Top: daily mean temperature (dashed line: threshold temperature $T_T = 0\,^\circ$C). Middle: daily precipitation (bars) and simulated snow storage (snow water equivalent, shaded). Bottom: monthly $\SPI{3}$ and $\SMRI{3}$ (dotted line: drought threshold $-1$).}
\label{fig:winter}
\end{figure}

The seasonal pattern is robust to the snow-model parameters (Supplementary Fig.~S1). With the fixed parameters, the monthly differences lie close to the median of the 200 parameter sets and within their 5--95\% range in every month, at every station and at both forecast accumulation periods. The parameters affect mainly the size of the differences in the snow season: at the 3-month accumulation period, the 95th percentile reaches 0.54 at Youngstown, 0.49 at Cleveland, and 0.34 at Columbus, about twice the largest value obtained with the fixed parameters, whereas the range is zero or nearly zero from July to October. Figure~\ref{fig:winter} illustrates the mechanism for the winter of 1977--78, which has the largest input-only differences averaged over the three stations. Precipitation stored in the snowpack in January and February lowered the February $\SMRI{3}$ well below the $\SPI{3}$ ($-0.74$ vs.\ 0.52 at Youngstown, $-1.06$ vs.\ 0.03 at Cleveland, and $-0.63$ vs.\ 0.91 at Columbus), and the melt in March raised it later: in April 1978 the $\SPI{3}$ indicated moderate drought at all three stations ($-1.76$, $-1.61$, and $-1.21$), whereas the $\SMRI{3}$ did not ($-0.76$, $-0.85$, and 0.02).

Such cases are frequent enough to change the drought classification. At the 3-month accumulation period, 25\% of the December--May months in drought (index below $-1$) according to either index at Youngstown are in drought according to only one of them, 32\% at Cleveland, and 21\% at Columbus (19, 23, and 14 months over 74 years), compared with at most 3\% from June to November (Supplementary Table~S6). Disagreements occur in both directions in similar numbers, so the snow adjustment changes \emph{when} drought is indicated rather than how often. The disagreement is smaller at longer accumulation periods: 8--11\% of the December--May drought months at 6 months, 9--14\% at 9 months, and at most 6\% at 12 months.

\subsection{Multiscale Differences Between Drought Representations}
\label{sec:res_wavelet}

The snow adjustment affects mainly the finest wavelet levels. Level by level, the SPI and SMRI decompositions differ mainly at the finest scales (Supplementary Table~S7). The correlation between corresponding levels is 0.91--0.96 for $L_1$, 0.96--0.98 for $L_2$, and 0.98--0.99 for $L_3$, and at least 0.99 for $L_4$--$L_6$ and the smooth component $S_6$; for $L_1$--$L_4$, the lowest correlations occur at Youngstown and Cleveland and the highest at Columbus. The differences in the finest level are also seasonal: at the 3-month accumulation period, the mean absolute difference in $L_1$ is largest from January to May (up to 0.15 at Youngstown, 0.13 at Cleveland, and 0.09 at Columbus) and at most 0.06 from June to October. The distribution of energy across levels is almost the same for the two indices. The share of $L_1$ is slightly higher for the SMRI, by 0.6 to 2.7 percentage points, and the shares of the other components differ by less than 1.3 percentage points. The snow adjustment thus adds a small amount of high-frequency variation in winter and spring and leaves the coarse-scale structure of the index essentially unchanged.

\subsection{Association Patterns and Wavelet-Level Rankings}
\label{sec:res_assoc}

Significance testing cannot discriminate between the wavelet levels. In the full-record decompositions, Kendall's $\tau$ between an index and its wavelet levels ranges from 0.22 to 0.55 (Supplementary Table~S8). All six levels are significant at the 5\% level for every index, accumulation period, and station (largest $p$ value $4\times10^{-23}$), and all 95\% moving-block bootstrap intervals exclude zero for block lengths of 6, 12, and 24 months. The interval widths increase with the block length, from 0.05--0.17 for 6-month blocks to 0.05--0.21 for 24-month blocks, but the lower limits are at least 0.13. The significance rule of \citet{evkaya2021} therefore retains every level and cannot be used to select components, which supports the prediction-driven selection of Section~\ref{sec:selection}.

The training-period rankings used to form the Top-$m$ sets are given in Supplementary Table~S9. At the 3-month accumulation period, $L_2$ and $L_3$ are ranked first and second at all stations, and at the 6-month accumulation period $L_3$ and $L_4$; the finest level $L_1$ ranks third or fourth at 3 months and last at 6 months. The rankings of the SPI and SMRI are identical at Columbus and for the 6-month indices at Cleveland. Elsewhere, the SPI and SMRI differ only in the order of the two highest-ranked levels, whose $|\tau|$ values differ by at most 0.02. The association with the index therefore does not distinguish the two representations; whether their wavelet levels carry different predictive information is examined with the forecasting experiment below.

\subsection{Model Selection on the Validation Period}
\label{sec:res_validation}

The validation period favored the wavelet inputs for the 3-month but not for the 6-month indices (Supplementary Table~S10; the selected inputs are listed in Table~\ref{tab:forecast}). For all six 3-month series, the best NARX specification had a lower median validation RMSE than NAR, by 0.02--0.05. The selected sets contained only the two highest-ranked levels, $L_2$ alone or $L_2$ and $L_3$. For all six 6-month series, NAR had the lower median validation RMSE, by 0.006--0.013; the best NARX specification would have used a single level ($L_3$ or $L_4$). Four hidden neurons were selected for every model except NAR for the Cleveland $\SMRI{6}$ ($H = 5$). For the linear benchmarks, the Akaike criterion on the training period selected $p = 10$--11 lags for the 3-month and $p = 7$--12 for the 6-month indices. Wavelet levels were selected for AR($p$)+W in eight of the twelve series, but they reduced the validation RMSE by at most 0.005, except for the Youngstown $\SMRI{6}$ (0.017).

\begin{table}[!htbp]
\centering
\caption{Selected inputs and test-period RMSE of the ensemble and reference forecasts. NARX levels: wavelet levels of the selected NARX model. AR($p$)+W: lag order chosen on the training period and wavelet set chosen on the validation period (``none'': no level selected, so that AR($p$)+W equals AR($p$)). NAR and NARX: RMSE of the ensemble forecast (monthly median of the 100 repetitions). The lowest RMSE of each series is shown in bold.}
\label{tab:forecast}
\footnotesize
\setlength{\tabcolsep}{3.5pt}
\begin{tabular}{llllrrrrr}
\toprule
& & \multicolumn{2}{c}{Selected inputs} & \multicolumn{5}{c}{Test RMSE} \\
\cmidrule(lr){3-4}\cmidrule(lr){5-9}
Station & Series & NARX levels & AR($p$)+W & AR(2) & NAR & NARX & AR($p$) & AR($p$)+W \\
\midrule
Youngstown & $\mathrm{SPI}\mbox{-}3$ & $L_3{+}L_2$ & $p=10$, none & 0.713 & 0.719 & 0.628 & \textbf{0.595} & \textbf{0.595} \\
 & $\mathrm{SMRI}\mbox{-}3$ & $L_2$ & $p=10$, none & 0.721 & 0.731 & 0.675 & \textbf{0.628} & \textbf{0.628} \\
 & $\mathrm{SPI}\mbox{-}6$ & $L_3$ & $p=7$, none & 0.506 & 0.508 & 0.511 & \textbf{0.476} & \textbf{0.476} \\
 & $\mathrm{SMRI}\mbox{-}6$ & $L_4$ & $p=12$, Top-5 & 0.518 & 0.519 & 0.522 & 0.503 & \textbf{0.471} \\
\addlinespace
Cleveland & $\mathrm{SPI}\mbox{-}3$ & $L_3{+}L_2$ & $p=11$, Top-6 & 0.696 & 0.694 & 0.625 & 0.604 & \textbf{0.598} \\
 & $\mathrm{SMRI}\mbox{-}3$ & $L_2{+}L_3$ & $p=11$, Top-6 & 0.715 & 0.712 & 0.647 & 0.634 & \textbf{0.627} \\
 & $\mathrm{SPI}\mbox{-}6$ & $L_4$ & $p=8$, none & 0.503 & 0.501 & 0.497 & \textbf{0.456} & \textbf{0.456} \\
 & $\mathrm{SMRI}\mbox{-}6$ & $L_4$ & $p=7$, Top-1 & 0.512 & 0.507 & 0.507 & \textbf{0.473} & \textbf{0.473} \\
\addlinespace
Columbus & $\mathrm{SPI}\mbox{-}3$ & $L_2$ & $p=10$, Top-4 & 0.688 & 0.708 & 0.631 & 0.588 & \textbf{0.585} \\
 & $\mathrm{SMRI}\mbox{-}3$ & $L_2$ & $p=10$, Top-4 & 0.705 & 0.720 & 0.650 & 0.613 & \textbf{0.609} \\
 & $\mathrm{SPI}\mbox{-}6$ & $L_3$ & $p=8$, Top-2 & 0.515 & 0.517 & 0.503 & 0.474 & \textbf{0.471} \\
 & $\mathrm{SMRI}\mbox{-}6$ & $L_3$ & $p=8$, Top-2 & 0.520 & 0.519 & 0.502 & 0.483 & \textbf{0.479} \\
\bottomrule
\end{tabular}

\end{table}

\subsection{Out-of-Sample Performance of NAR and NARX}
\label{sec:res_test}

\begin{figure}[!htbp]
\centering
\includegraphics[width=\textwidth]{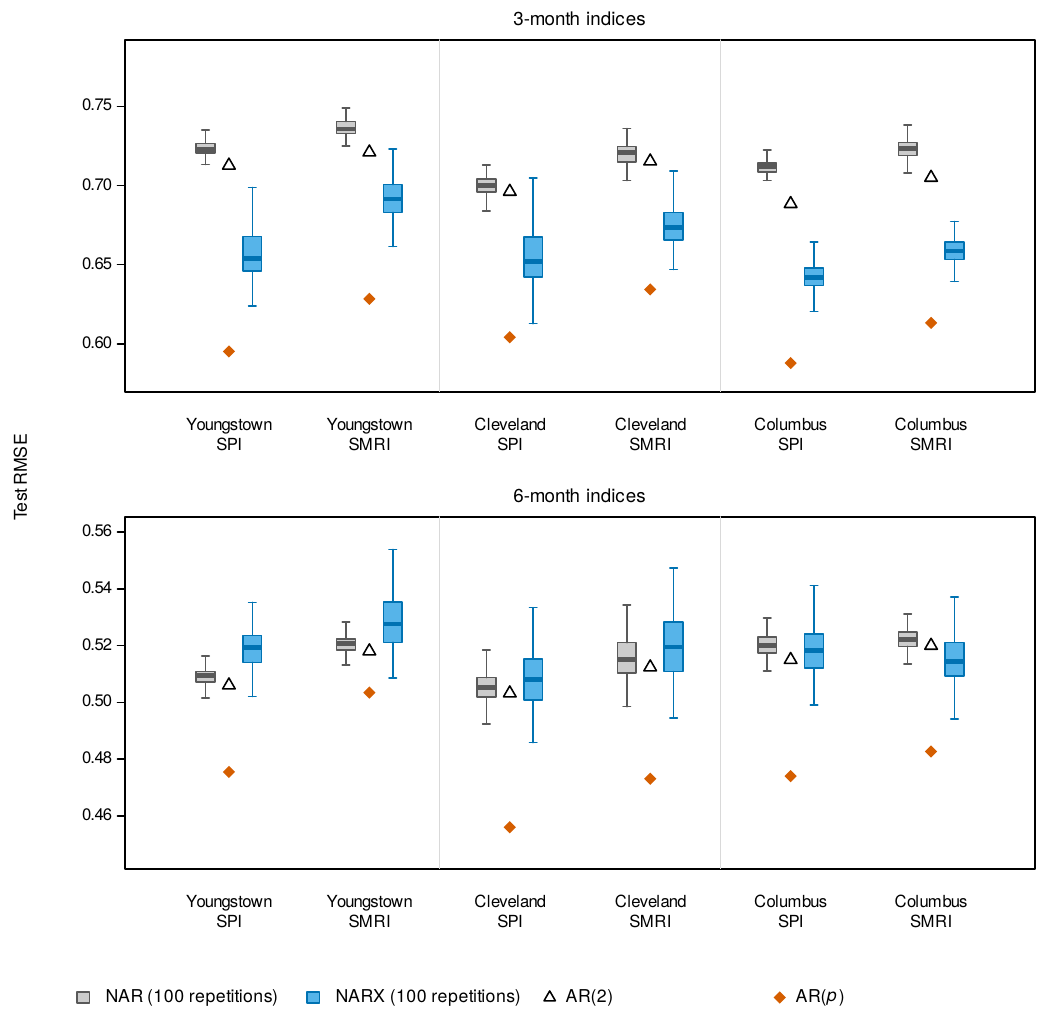}
\caption{Test-period RMSE of NAR and NARX over 100 paired repetitions (boxes: interquartile range; whiskers: 1.5 times the interquartile range; repetitions beyond the whiskers, including failed fits, are not shown) and of the AR(2) and AR($p$) benchmarks, for the 3-month (top) and 6-month (bottom) indices.}
\label{fig:forecast}
\end{figure}

On the test period, the wavelet inputs improved the 3-month forecasts consistently and the 6-month forecasts not at all, as the validation period had indicated (Fig.~\ref{fig:forecast}; Supplementary Table~S11). For the 3-month indices, NARX reduced the median RMSE of NAR by 6.3--9.7\% (MAE by 7.9--14.4\%), and it was more accurate than NAR in 92--100\% of the paired repetitions at every station. The difference was significant at the 5\% level in 35--95\% of the individual repetitions, and the NARX ensemble was significantly more accurate than the NAR ensemble for all six 3-month series ($p \le 0.003$; Supplementary Table~S12). For the 6-month indices, the change in the median RMSE ranged from $-2.1$ to $+1.6$\%, NARX was better in 9--82\% of the repetitions, and the differences were not significant (ensemble $p \ge 0.10$; significant in at most 4\% of the repetitions). The variability caused by the random starting weights was small compared with these differences: at most two of the 100 repetitions of any model had an RMSE above 1.

The gain from the wavelet inputs was of similar size for the two drought representations. At the 3-month accumulation period, it was 9.5\% for the SPI and 6.3\% for the SMRI at Youngstown, 6.8\% and 6.3\% at Cleveland, and 9.7\% and 8.9\% at Columbus. The snow-aware representation therefore did not make the index more predictable from its wavelet levels.

\subsection{Comparison with Reference Forecasts}
\label{sec:res_benchmarks}

A linear autoregression with a longer memory was at least as accurate as the two-lag NARX for every series (Table~\ref{tab:forecast} and Fig.~\ref{fig:forecast}; the full set of measures is given in Supplementary Table~S13, and the Diebold--Mariano tests in Supplementary Table~S12). Climatology, with an RMSE of 0.97--1.06, and persistence, with 0.72--0.80 for the 3-month and 0.53--0.54 for the 6-month indices, were the least accurate. The NAR ensemble was not significantly more accurate than the linear AR(2) model with the same two lags at any station, and at Columbus AR(2) was significantly better for both 3-month indices ($p = 0.002$ and 0.006). For the 3-month indices, the NARX ensemble was significantly more accurate than persistence ($p \le 0.003$) and AR(2) ($p < 0.02$) at all stations.

The autoregression with a longer memory, however, was more accurate still. AR($p$) had a lower test RMSE than the two-lag NARX ensemble for all twelve series: 0.588--0.634 compared with 0.625--0.675 for the 3-month indices, and 0.456--0.503 compared with 0.497--0.522 for the 6-month indices. The difference was significant in six series (Youngstown $\SPI{3}$, $\SMRI{3}$, and $\SPI{6}$; Columbus $\SPI{3}$, $\SMRI{3}$, and $\SPI{6}$; $p = 0.017$--0.041) and not significant in the others. Adding the selected wavelet levels to AR($p$) changed the RMSE by at most 0.008, except for the Youngstown $\SMRI{6}$, where it fell from 0.503 to 0.471 ($p = 0.007$); for the other series in which levels were selected, the differences were not significant ($p \ge 0.26$).

Networks with more lags confirm that memory is the main factor (Supplementary Table~S14). For the 3-month indices, the validation period selected six lags for all six series (median validation RMSE 0.594--0.687, against 0.640--0.723 with two lags). With six lags, the median test RMSE of NAR fell from 0.700--0.736 to 0.633--0.688, and the gain of NARX over NAR decreased from 6.3--9.7\% to between $-0.8$ and 4.5\%, with NARX more accurate in only 39--84\% of the paired repetitions. The NARX ensemble was then about as accurate as AR($p$): its RMSE was 0.595--0.662, compared with 0.588--0.634 for AR($p$), and it was lower than that of AR($p$) at Cleveland for both indices. The difference was significant only for the Youngstown $\SMRI{3}$ (0.662 vs.\ 0.628, $p = 0.018$; $p \ge 0.23$ for the other series).

\subsection{Seasonal Performance and Drought Events}
\label{sec:res_season_fc}

\begin{figure}[!htbp]
\centering
\includegraphics[width=\textwidth]{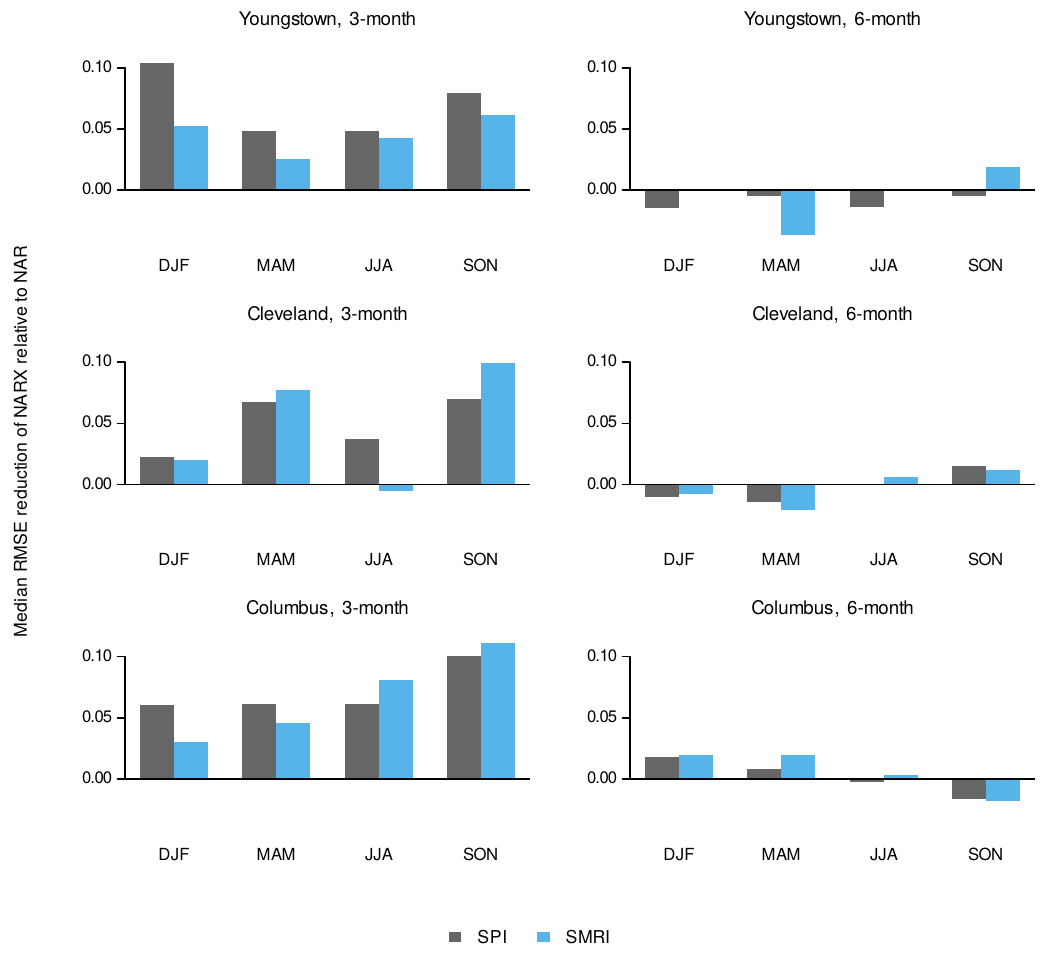}
\caption{Median reduction of the test-period RMSE of NARX relative to NAR over the paired repetitions, by meteorological season (DJF: December--February; MAM: March--May; JJA: June--August; SON: September--November), for the SPI (gray) and the SMRI (blue).}
\label{fig:seasonal_gain}
\end{figure}

The NARX gain was not concentrated in the snow season, and it was not larger for the SMRI. For the 3-month indices, it was positive in all four seasons at every station, with one exception (Cleveland $\SMRI{3}$ in summer, $-0.005$), and it was largest in autumn at Cleveland and Columbus (0.07--0.11) and in winter for the Youngstown $\SPI{3}$ (0.10) (Fig.~\ref{fig:seasonal_gain}). It was not larger for the SMRI than for the SPI in winter and spring, the seasons in which the two indices differ most; the only exception was Cleveland in spring (0.077 vs.\ 0.068). For the 6-month indices, the seasonal gains lay between $-0.04$ and $0.02$.

The test period contained 11--16 drought months (index below $-1$) for the 3-month and 4--13 for the 6-month series. At a lead time of one month, all models missed most of them: over the twelve series, the median probability of detection was 0.39 for persistence (with a median false alarm ratio of 0.61), 0.33 for AR($p$) (0.43), 0.20 for the NARX ensemble (0.50), and 0.19 for the NAR ensemble (0.38) (Supplementary Table~S13). With so few events, these differences are not robust, and none of the models provided a reliable warning of the onset of drought one month ahead.

\section{Discussion}
\label{sec:discussion}

The first question was whether accounting for snow changes a drought index in a region where snow is seasonal rather than dominant. Over the whole record the SPI and SMRI are almost interchangeable, but the overall correlations hide a clear seasonal signal. At the 3-month accumulation period, the two indices differ most from December to May, the differences are attributable to the snow-adjusted input rather than to the fitted distribution, and their seasonal pattern does not depend on the snow-model parameters. The size of the mean difference follows the snow climate of the stations (Table~\ref{tab:stations}): it is largest at Youngstown, the coldest station, and smallest at Columbus, where only about 5\% of the precipitation falls on days below freezing. The practical consequence is a change in drought classification: in winter and spring, one in five to one in three drought months at the 3-month accumulation period is identified by only one of the two indices. The winter of 1977--78 is an example: an SPI-based assessment would have indicated moderate drought in April at all three stations, although much of the winter precipitation had only just reached the ground as melt. These results agree with the findings of \citet{staudinger2014} for snow-dominated catchments, and they show that the timing effect of snow is also visible in a humid, low-elevation region where snow cover is intermittent \citep{suriano2024,ford2021}. For drought monitoring in such regions, the SMRI can therefore complement the SPI in winter and spring at short accumulation periods, which are also the periods most relevant for soil moisture and early-season agriculture.

The effect of snow weakens as the accumulation period increases, because the delay between snowfall and melt is short compared with a 6- or 12-month window. At these longer periods, the remaining differences between the SPI and SMRI are of the same order as those caused by the choice between the Gamma and Pearson type~III distributions, and in some spring months at Cleveland and Columbus they are mainly attributable to the distribution. Studies that compare a snow-aware and a precipitation-based index fitted with different distributions should therefore separate the two effects, as done here with the auxiliary indices, before attributing differences to snow.

The second question concerned forecasting. In a leakage-free design, the wavelet inputs improved the one-month-ahead forecasts of the 3-month indices consistently, at all stations and for almost all random initializations, whereas they did not improve the 6-month forecasts, and the validation period identified these cases correctly. This result agrees with the conclusion of \citet{evkaya2021} that wavelet series can improve NARX forecasts of the SPI, and it shows that the improvement does not depend on future information. However, the third question shows that the gain has a simpler explanation. An autoregression with a lag order of 7--12 months, chosen on the training period, was at least as accurate as the two-lag NARX for every series, and adding wavelet levels to it rarely helped. The NAR and NARX models of the original framework use only the last two values of the index. A $k$-month index is a transformed moving sum of $k$ monthly values and is autocorrelated over many lags, so that its next value depends on information that two lags do not contain, in particular on the monthly value that will leave the accumulation window. The causal wavelet levels supply part of this information because they summarize a longer history. In the linear models, adding the two highest-ranked levels ($L_2$ and $L_3$), which were also the levels preferred by the training ranking and by NARX, did not change the forecasts: with the reflection boundary rule, their end values are linear combinations of the recent values of the index, which the autoregression already contains. The improvement attributed to the multiscale inputs is therefore largely an improvement in memory. Networks with longer memory support this interpretation: when six index lags were allowed, the gain of the wavelet levels fell to between $-0.8$ and 4.5\% and the networks reached the accuracy of the autoregression.

This finding complements the methodological concerns raised by \citet{quilty2018} and \citet{du2017}. Those studies showed that wavelet-based forecasts can be overly optimistic when future information enters the inputs. The present results show that, even when the inputs are constructed correctly, the gain of a wavelet hybrid can be overstated if the benchmark has a shorter memory than the hybrid. Wavelet--neural-network models for accumulated drought indices should therefore be compared not only with the corresponding single network but also with a linear autoregression whose order is chosen on the training data. The Diebold--Mariano test and paired repetitions, as used here, make such comparisons more reliable than single error values.

The snow-aware representation did not change these conclusions. The gains of NARX over NAR, the selected wavelet levels, and the ranking of the models were similar for the SPI and the SMRI, and the gain was not larger in winter and spring. Snow thus changes what the index says about water availability in these seasons, but it does not make the index easier or harder to forecast one month ahead from its own history. Whether the snowpack itself, for example the simulated snow storage at the forecast origin, can improve forecasts of the SMRI is a different question that was not examined here.

Several limitations should be noted. The study covers three stations in one state, and the conclusions may differ in regions with deeper or more persistent snow cover. The snow model was not calibrated, and its output was not compared with observed snow depth or snowfall; the sensitivity analysis shows that the seasonal pattern of the SPI--SMRI differences is robust to the parameters, but the size of the winter differences is not. The forecasting experiment uses a single chronological split, so that all conclusions refer to the test period from 2011--2012 to 2026, and it considers one-month-ahead forecasts only. The main networks follow the configuration of \citet{evkaya2021}, with two lags and four to six hidden neurons; networks with longer lags were examined only for the 3-month indices and with four hidden neurons. Finally, the indices are standardized with distributions fitted to the whole study period, as is common for the SPI, and the test period contains few drought months, so that the drought-event scores are only indicative.

\section{Conclusion}
\label{sec:conclusion}

This study compared the SPI and the snow-aware SMRI at three stations in Ohio and examined whether their wavelet components improve one-month-ahead drought forecasts in a leakage-free design. Accounting for snow changes the 3-month drought index mainly from December to May, by up to about a quarter of a standard deviation on average in the snowiest months; the change is attributable to the delayed water input rather than to the fitted distribution, it is robust to the snow-model parameters, it follows the snow climate of the stations, it changes the drought classification of 21--32\% of the December--May drought months, and it weakens at longer accumulation periods. In forecasting, causal wavelet inputs improved two-lag NAR networks by 6--10\% for the 3-month indices, consistently across stations and random initializations, but not for the 6-month indices. A linear autoregression with a longer memory, chosen on the training period, was at least as accurate, and wavelet levels added little to it. When the networks were given six index lags, the wavelet gain fell to at most 4.5\% and they matched the autoregression. The forecast gain of the wavelet inputs therefore reflects the memory of the accumulated index rather than additional multiscale information, and it is similar for the SPI and the SMRI. For regions with seasonal snow, the SMRI is a useful complement to the SPI for characterizing short-term drought in winter and spring, and wavelet-based drought forecasts should be judged against linear benchmarks with an appropriate memory.

\section*{Declarations}

\bmhead{Funding}
This research received no specific grant from any funding agency in the public, commercial, or not-for-profit sectors.

\bmhead{Competing interests}
The authors declare that they have no conflict of interest.

\bmhead{Data availability}
The daily GHCN-Daily records used in this study are publicly available from the NOAA National Centers for Environmental Information (Climate Data Online) for stations USW00014852, USW00014820, and USW00014821. The files used, as downloaded, are provided with the code.

\bmhead{Code availability}
The R package and analysis scripts that reproduce all results, tables, and figures, together with the raw daily GHCN-Daily data used, are provided as online supplementary material with this submission. All intermediate and final results, including those not reproduced in the Supplementary Information, are included there as CSV files.

\bmhead{Author contributions}
Symon Forthman and Max Gluck contributed equally to this work and are listed alphabetically: Data curation, Investigation, Validation. 
Fatma Sevin\c{c} Kurnaz: Conceptualization, Methodology, Software, Formal analysis, Data curation, Writing -- original draft, Writing -- review \& editing, Supervision, Project administration. All authors read and approved the final manuscript.

\bmhead{Acknowledgements}
The authors thank Case Western Reserve University for institutional support and acknowledge the NOAA National Centers for Environmental Information for maintaining the GHCN-Daily database used in this study.

\bmhead{Declaration of generative AI and AI-assisted technologies in the writing process}
During the preparation of this work, the authors used a large language model (Claude, Anthropic) to review and edit the authors' analysis code and manuscript text. The authors wrote the code and the manuscript, designed the study, specified every methodological decision, and verified the code and the results. The authors take full responsibility for the content of the published article.

\bibliography{references}

\end{document}